\documentclass[showpacs,preprintnumbers,amsmath,amssymb]{revtex4}
\draft
\usepackage{graphicx}
\begin{document}
\def \tr{{\mbox{tr~}}}
\def \ra{{\rightarrow}}
\def \ua{{\uparrow}}
\def \da{{\downarrow}}
\def \be{\begin{equation}}
\def \ee{\end{equation}}
\def \bea{\begin{eqnarray}}
\def \eea{\end{eqnarray}}
\def \nn{\nonumber}
\def \half{{1\over 2}}
\def \etal{{\it {et al}}}
\def \cH{{\cal{H}}}
\def \cF{{\cal F }}
\def \cT{{\cal T }}
\def \cM{{\cal{M}}}
\def \cN{{\cal{N}}}
\def \cQ{{\cal Q}}
\def \cI{{\cal I}}
\def \cV{{\cal V}}
\def \cG{{\cal G}}
\def \bS{{\bf S}}
\def \bL{{\bf L}}
\def \bG{{\bf G}}
\def \bQ{{\bf Q}}
\def \bR{{\bf R}}
\def \br{{\bf r}}
\def \bu{{\bf u}}
\def \bq{{\bf q}}
\def \bk{{\bf k}}
\def \bPsi{{\bf \Psi}}
\def \bpsi{{\bf \psi}}
\def \tJ{{\tilde{J}}}
\def \W{{\Omega}}
\def \e{{\epsilon}}
\def \lam{{\lambda}}
\def \a{{\alpha}}
\def \t{{\theta}}
\def \b{{\beta}}
\def \g{{\gamma}}
\def \D{{\Delta}}
\def \d{{\delta}}
\def \w{{\omega}}
\def \s{{\sigma}}
\def \f{{\varphi}}
\def \x{{\chi}}
\def \h{{\eta}}
\def \hatt{{\hat{\t}}}
\def \tt{{\tilde{t}}}
\def \hn{{\bar{n}}}
\def \vk{{\bf{k}}}
\def \vq{{\bf{q}}}
\def \gk{{\g_{\vk}}}
\def \nd{{^{\vphantom{\dagger}}}}
\def \yd{^\dagger}
\def \ket#1{{\,|\,#1\,\rangle\,}}
\def \bra#1{{\,\langle\,#1\,|\,}}
\def \braket#1#2{{\,\langle\,#1\,|\,#2\,\rangle\,}}
\def \expect#1#2#3{{\,\langle\,#1\,|\,#2\,|\,#3\,\rangle\,}}
\def \rl#1#2{{\,\langle\,#1\,#2\,\rangle\,}}
\def \ad{{\a\yd}}
\def \an{{\a\nd}}
\def \av#1{{\langle#1\rangle}}
\def \bd#1{{(\sin\t\ad_{0#1}+\cos\t\ad_{1#1})}}
\def \bn#1{{(\sin\t\an_{0#1}+\cos\t\an_{1#1})}}
\def \sd#1{{(\cos\t\ad_{0#1}-\sin\t\ad_{1#1})}}
\def \sn#1{{(\cos\t\an_{0#1}-\sin\t\an_{1#1})}}
\draft
%\twocolumn[\hsize\textwidth\columnwidth\hsize\csname @twocolumnfalse\endcsname

\title{Phase diagram of two--component bosons on an optical lattice.}
\author{Ehud Altman, Walter Hofstetter, Eugene Demler and Mikhail D. Lukin}
\affiliation{Department of Physics, Harvard University}
\date{\today}
\begin{abstract}
We present a theoretical analysis of the phase diagram
of two--component bosons on an optical lattice. A new formalism is developed
which treats the effective spin interactions in the Mott and superfluid phases on the same
footing. Using the new approach we chart the phase boundaries of the broken spin symmetry states
up to the Mott to superfluid transition and beyond.
Near the transition point, the magnitude of spin exchange
can be very large, which facilitates the experimental realization
of spin-ordered states. We find that spin and quantum
fluctuations have a dramatic effect on the transition making it first order in extended regions of
the phase diagram. For Mott states with even occupation we find that
the competition between effective Heisenberg exchange
and spin-dependent on--site interaction leads to an additional
phase transition from a Mott insulator with no broken symmetries into a
spin--ordered insulator.
\end{abstract}
\pacs{PACS} \maketitle
%\vskip2pc]
%\narrowtext

\section{Introduction}
Recent observations of the superfluid to Mott insulator transition
in a system of ultracold atoms in an optical lattice open
fascinating prospects for studying many--body phenomena associated
with strongly correlated systems in a highly controllable
environment \cite{greiner,orzel,jaksch}. For instance, theoretical
studies have shown that, with spinor bosonic or fermionic atoms in
optical lattices, it may be possible to observe complex quantum
phase transitions \cite{demler-zhou}, to realize novel
superfluidity mechanisms \cite{walter}, and to probe
one--dimensional systems exhibiting spin charge separation
\cite{recati}.

%%%%%%%%%%%%%%%%%%%%%%%%%%%%%%%%%%%%%%%%%%%%%%%%%%%%%
Recently, Duan {\it et al.} \cite{duan} proposed a technique to implement interacting
spin-$\half$ Hamiltonians using ultra-cold atoms,
opening the door to controlled studies of quantum magnetism. In
this approach the two-state bosonic or fermionic atoms are
confined in an optical lattice where spin--dependent interactions and
hopping  are controlled by adjusting the intensity, frequency, and
polarization of the trapping light. Deep in the Mott phase the motional
degrees of freedom are frozen out and the remaining spin degrees of
freedom are coupled by an effective Heisenberg exchange. In refs. \cite{duan,kuklov},
an effective
spin hamiltonian was derived by perturbation theory
for the case of a single atom per site and the limit of small tunneling.
However, in practice Mott states with more than one atom per site
are also of considerable interest
and may exhibit richer phase diagrams. Furthermore, spin effects are expected
to be important, and even stronger, at larger values of the
tunnelling, where perturbation theory fails. For example, an important question
that cannot be addressed by the perturbative treatments is how
spin affects the transition into a superfluid phase and the
properties of the superfluid phase itself.

In this paper we first extend the earlier approaches to the case of Mott
states with general integer occupation. We find that at even
fillings the competition between on-site interactions and nearest
neighbor spin exchange leads to a transition from a spin ordered
Mott state to one with no broken symmetries. Then we present a
theoretical framework, which is non--perturbative in the
tunneling and allows to describe both the superfluid and
insulating phases in two-component systems. Using this approach we
determine the phase diagram for a density of one atom per
site. We find that the spin--ordered states persist up to the
superfluid transition. In this region the critical temperature for
spin ordering can be large, facilitating experimental realization
of these phases. The z-antiferromagnetic state in particular,
enjoys a negative zero-point energy which extends its domain far
beyond the mean field prediction for the Mott phase. The
transition between this state and the superfluid is found to be
first order in contrast with the standard superfluid-insulator
transition.

Before proceeding, we note that spin Hamiltonians can also be
simulated by controlled collisions via frequent time-dependent
shifts of the lattice potentials \cite{lloyd}. Compared with that
method, the spin-dependent tunneling may have certain
experimental advantages since it implements the desired
Hamiltonian directly, thus circumventing imperfections and errors
associated with rapid perturbations due to the lattice shifts. We
also note the recent studies on quantum magnetism induced via
magnetic dipole interactions of the condensed atoms
\cite{meystre}. The present approach results in much larger
interaction strength per atom, and also allows for more
flexible control over interaction properties.

The paper is organized as follows. In section \ref{sec:BHM} we describe the Hubbard model
for two bosonic species on an optical lattice, which serves as our starting point.
In section \ref{sec:Natom}, the perturbative approach of refs. \cite{duan,kuklov} is extended to
arbitrary integer filling and the insulating phase diagram is investigated.
In section \ref{sec:MFT}
we present the mean-field description of the SF-MI transition in two component systems.
The analytical predictions of a variational approach
are compared to the results of a numerical mean-field
analysis.
In section \ref{sec:fluc} a theoretical framework is developed that incorporates
the effect of quantum fluctuations and treats the magnetic interactions
in the Mott and superfluid phases on an equal footing.
In Section \ref{sec:phase-diagram},
this framework is  used to analyze the full phase diagram for one atom per
lattice site. The relevance of the present results in the light of realistic
experiments is discussed in the concluding paragraphs of the paper.

\section{The model}\label{sec:BHM}

We consider a system with two species of atoms
or equivalently, atoms with two relevant internal states.
The two species shall be denoted by the second quantized bosonic operators
$a$ and $b$. We assume that the two species are trapped by independent standing
wave laser beams through polarization (or frequency) selection. Each
laser beam creates a periodic potential in a certain direction
$v_{\a\sigma} sin^2(\vec{k}_\a \vec{r})$, where $\vec{k}_\a$
is the wavevector of the light and $\s=a,b$ is the species index.
Throughout this work, we assume that the laser beams are orthogonal,
creating either a square lattice in two dimensions or a cubic lattice
in three dimensions.
For sufficiently strong periodic potential and
low temperatures the atoms will be confined to
the lowest Bloch band. The low--energy  Hamiltonian is then given by
the Bose-Hubbard model for two boson species:
\bea
H&=&-\sum_{\av{ij}}t_a\left(a\yd_i a\nd_j +H.c\right)
-t_b\sum_{\av{ij}}\left(b\yd_i b\nd_j+H.c\right)\nn\\
&&+U\sum_i(n_{ai}-\half) (n_{bi}-\half)\nn\\
&&+\half\sum_{i \a=a,b}V_\a n_{\a i}(n_{\a i}-1)
-\sum_{i \a}\mu_\a n_{\a i}
\label{BHM}
\eea
Here $\left\langle i,j\right\rangle $ denotes the near neighbor sites,
$a_{i},b_i$ are bosonic annihilation
operators respectively for bosonic atoms of different spin states
localized on site $i$, $n_{i a }=a_{i_{\sigma }}^{\dagger
}a_{i_{\sigma }}$, $n_{i b}=b_{i_{\sigma }}^{\dagger
}b_{i_{\sigma }}$.
For the cubic lattice,  using a harmonic approximation
around the minima of the potential \cite{jaksch}, the spin-dependent tunneling
energies and the on-site interaction energies are given by $t_{a(b)}
\approx \left( \pi ^{2}/4\right) v_{a(b) }\exp [ -\left( \pi
^{2}/4\right) (v_{a(b) }/E_{R})^{1/2}]$, $U\approx (8/\pi)^{1/2} ( ka_{a,b }) ( E_{R}\overline{v}
_{ab }^3) ^{1/4}$. Here $v_{a,b}$ is the depth of the optical potential for species $a$ and $b$,
$\overline{v}_{ab }=4v_{a }v_{b }/( v_{a}^{1/2}+v_{b }^{1/2}) ^{2}$ is the spin average potential in
each direction, $E_{R}=\hbar ^{2}k^{2}/2m$ is the atomic recoil energy, and $%
a_{ab }$ is the scattering length between the atoms of
different spins. The intra--species interaction is given by
$V_{a(b)}\approx (8/\pi)^{1/2} \left(ka_{a(b) }\right) \left( E_{R}v_{a(b) }^3\right)
^{1/4}$ ($a_{a(b) }$ are the corresponding scattering lengths). Furthermore,
the magnitude of the interspecies interaction $U$ can be additionally controlled
by shifting the two lattices away from each other,
which opens a wide range of $U/V_\a$ to exploration. Note
that spin-dependent tunnelling $t_{\mu \sigma }$ can be easily introduced by
varying the potential depth $v_{a }$ and $v_{b }$
with control of the intensity of the trapping laser. We should also point out that the
two atomic states generally have different energies ($\mu_a\ne\mu_b$ in \ref{BHM}).
In the spin language, this
translates to a magnetic field in the $z$ direction. However since there is essentially
no transfer between the two populations, the experiment is performed with fixed magnetization
and the chemical potentials can be set to fix this magnetization.

In this paper, we address primarily the case in which the total filling
is commensurate with the lattice and the two species have equal density.
A transition from a superfluid to a Mott
insulator is expected, as in the usual case of a single species.
However, in this system, magnetic order,
associated with the pseudospin degrees of freedom (boson components), may occur as well.

\section{Deep Mott phase: effective spin hamiltonian}\label{sec:Natom}

\begin{figure}[h]
  \centering
  \includegraphics[width=6cm]{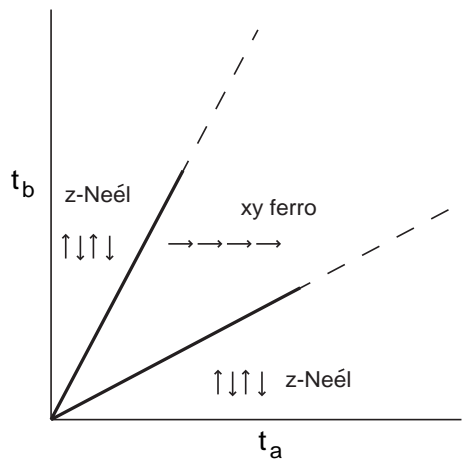}
\caption{Schematic phase diagram of the effective spin hamiltonian
\protect{(\ref{Hduan})}, valid deep in the Mott phase.}
\label{fig:pd-spin}
\end{figure}

To illustrate the magnetic orders that can arise it is instructive
to begin deep in the Mott insulator in the limit $t_{a,b}<<U,V_{a,b}$, where
the hamiltonian (\ref{BHM}) can be simplified considerably.
The low energy Hilbert space in this case contains states with a particular integer
occupation on every site. However there is a remaining degeneracy associated with the
spin (boson component) degrees of freedom.
The degeneracy can be removed by an effective hamiltonian acting within the
low energy subspace.
This was done previously in Refs \cite{kuklov,duan} for the case of a {\em single atom per site},
using second order perturbation theory in the hopping parameters.
The result is
\be
H_{eff}=J_z\sum_{\av{ij}}S^z_i S^z_j-J_\perp\sum_{\av{ij}}
(S^x_i S^x_j+S^y_i S^y_j)-h\sum_iS^z_i.
\label{Hduan}
\ee
Here $\ket{\ua}$ and $\ket{\da}$ represent sites occupied
by the $a$ and $b$ atoms respectively and the couplings are given by:
\bea
J_z&=&2\frac{t_b^2+t_a^2}{U}-\frac{4t_a^2}{V_a}-\frac{4t_b^2}{V_b}\nn\\
J_\perp&=&\frac{4t_a t_b}{U}\nn\\
h&=&\frac{2t_a^2}{V_a}-\frac{2t_b^2}{V_b}+h_{ext}
\eea
We assume that the induced ordering field $h$ can be cancelled
by an externally applied field $h_{ext}$. In this case the
model obviously exhibits a transition between a
$x-y$ ferromagnet for $J_{\perp}>J_z>0$ to an Ising antiferromagnet with
$z$-Neel order (Fig. \ref{fig:pd-spin}).
%%%%%%%%%%%%%%%%%%%%%%%%%%%%%%%%%%%

We now extend the discussion to the case of any integer
filling of $N$ atoms per site. To see how things may become qualitatively different
from the singly occupied case, consider first a Mott
state with two atoms per site. The low--energy Hilbert space of
a lattice site consists of the three states
\be
\frac{1}{\sqrt{2}}(a\yd)^2\ket{0},
~~a\yd b\yd\ket{0},~~\frac{1}{\sqrt{2}}(b\yd)^2\ket{0}.
\label{sb1}
\ee
If $V_{a,b}>>U$, the state $a\yd b\yd\ket{0}$ has much lower energy than the
other two. This implies a simple Mott state $\prod_i a\yd_i b\yd_i\ket{0}$ which,
unlike the Mott states in Fig. \ref{fig:pd-spin},
does not break any symmetries. On the other hand, when $V_{a,b}$ is of the same
order as $U$, all three states should be taken into account and more interesting
phases may be possible. Therefore in the general case of $N$ atoms per site
we consider the regime $t_{a,b},|V_{a,b}-U|<<U,V_{a,b}$ and $V_a=V_b$.

The low--energy Hilbert space of a lattice site with $N$ atoms per site can be constructed
in a similar way. It contains the $N+1$ states:
\be
\ket{S,m}=\frac{(a\yd)^{S+m}}{\sqrt{(S+m)!}}\frac{(b\yd)^{S-m}}{\sqrt{(S-m)!}}
\ket{0}
\label{sbN}
\ee
where $S\equiv N/2$ and $m=-S,\ldots,S$. Obviously $a\yd$ and $b\yd$ act as
Schwinger bosons creating a multiplet of pseudospin $S$. The spin magnitude
depends on the site occupancy. It is integer for even $N$ and half integer for odd $N$.

Now the effective hamiltonian within the spin $S$ subspace can be derived by second
order perturbation theory in a straightforward generalization of Ref. \cite{duan}.
The result is
\be
H_{eff}=-\sum_{\av{ij}}\left[J_\perp(S^x_i S^x_j+S^y_i S^y_j)+J_z
S^z_i S^z_j\right] +u\sum_i(S^z_i)^2-h\sum_i S^z_i
\label{HN}
\ee
where the interactions are given by:
\bea
u&=& V_a-U=V_b-U\nn\\
J_\perp&=& \frac{4t_a t_b}{U}\nn\\
J_z&=&2\frac{t_a^2+t_b^2}{U}\nn\\
h&=&z(2S+1)\frac{t_a^2-t_b^2}{U}+h_{ext}
\eea
Note that if we take $S\ra\half$ the parameters are identical to those
of the effective spin-$\half$
hamiltonian (\ref{Hduan}) in the case $U\approx V_a=V_b$. Note also that additional terms
such as $(S^z_i S^z_j)^2$ do not arise.
The $(S^z)^2$
term is of course just a constant for $S=\half$ and therefore has no effect in this
case. However,
it plays an important role at larger value of the spin, namely
for occupations $N>1$.

When $u$ can be neglected relative to $J_z$ or $J_\perp$, the remaining terms
in (\ref{HN}) form an anisotropic Heisenberg model.
A standard, coherent state mean field theory is then possible, which
yields $x-y$ ferromagnetic order.
A large positive $u$, in the absence of an ordering field ($h=0$),
acts to reduce the $S_z$ component of the spins. At some point,
the classical coherent states
that represent fully polarized spins, become unsuitable descriptions of the system.
In particular, at large enough $u$ all spins will be essentially confined to their lowest possible
$S^z$ states.

When the spin is half integer (odd filling), there are two
active states, at large $u$, corresponding to $S^z=\pm\half$. The hamiltonian (\ref{HN}) then
reduces to a spin-$\half$ model, but the spin interactions
remain practically identical. Thus, the
essential physics is
unchanged. We expect ferromagnetic spin order, as for small $u$,
only with a reduced effective moment.

For integer spin (even filling) we expect qualitatively different
behavior. At large enough $u$ only the $S^z=0$ state
will be important. We then expect that the system is well described by
$\ket{\Psi}=\prod_i\ket{S,m=0}_i$, a Mott state with no broken symmetries.
The transition to this state at large $u$ from the $x-y$ ferromagnet at small $u$
is formally identical to the transition from a superfluid to a Mott phase
in the Hubbard model of single component
bosons. A direct correspondence exists between the boson number in the Hubbard model and
$S^z$ in (\ref{HN}). The Mott state of bosons is characterized by
vanishing particle number fluctuations on a site. Similarly the transition here
is into a state with
well defined $S^z$ on each site.

\begin{figure}[h]
  \centering
  \includegraphics[width=6cm]{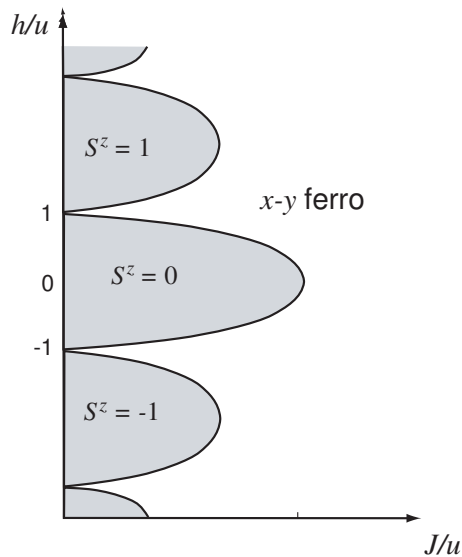}
\caption{Phase diagram of the Mott state with even number
of atoms per site. The lobes mark Mott states
with fixed $S^z=(n_a-n_b)/2$.
Outside of these lobes the system is also in a Mott state
but with $x-y$ ferromagnetic spin order.}
\label{fig:Natom-pd}
\end{figure}

To describe the transition we note that only the three
states with lowest $S^z$ play an important role in its vicinity. We therefore write a
homogenous mean field ansatz:
\be
\ket{\Psi} =\prod_i\Big[\cos (\t/2 )
\ket{S,0}+e^{i\h}\sin (\t/2 )
\left(e^{i\f}\cos (\x/2 ) \ket{S,1}
+e^{-i\f}\sin (\x/2 )
\ket{S,-1}\right)\Big].
\ee
The variational energy in this state is given by:
\bea
E=-\frac{J_\perp z}{8}S(S+1)\sin^2\t(1+\sin\x\cos 2\h)-\frac{J_z z}{2}sin^4\frac{\t}{2}\cos^2\x
+u\sin^2\frac{\t}{2}-h\sin^2\frac{\t}{2}\cos\x
\eea
and we see that the minimum occurs for $\h=0,\pi$.
The $x-y$ order parameter is $\av{S^+}\propto\sin\t\equiv\psi$. Therefore, to find the
transition to a Mott insulator we expand the energy up to quadratic order in $\psi$ and
minimize it with respect to $\x$.
Note that the quartic term is always positive since $J_z<J_\perp$. We then obtain the
critical value of $J_\perp$ as a function of $h$:
\be
\frac{J_{\perp c}}{u}=\frac{1-(h/u)^2}{z S(S+1)}
\ee
For magnetic fields $h>u$, the description in terms of the states $\{\ket{-1},\ket{0},\ket{1}\}$
breaks down. Instead, a similar scheme can be carried out,
using the states $\{\ket{0},\ket{1},\ket{2}\}$. This yields another lobe corresponding
to a phase with well defined $S^z=1$. A schematic phase diagram
is plotted in Fig \ref{fig:Natom-pd}. As the ordering field is increased, we obtain
lobes corresponding to larger values of $S^z$ up to $S^z=S$, where the spin is fully polarized.
In practice the number of particles in each spin state is conserved
independently. In other words the experiment is done with fixed $z$ magnetization
and $h$ is used as a theoretical tool to set this magnetization in our model.
Here we fix zero magnetiztion by setting $h=0$.

%%%%%%%%%%%%%%%%%%%%%%%%%%%%%%%%%%%

In summary, we found that the Mott phases of two--component
bosons with even site filling are markedly different from those with odd filling.
At odd filings, the Mott regions of the phase diagram are
essentially the same as those found for single occupation (see Fig. \ref{fig:pd-spin}).
These are all broken symmetry phases, either a $x-y$ ferromagnet or a $z$-Neel state.
The Mott phases at even filling are sketched in Fig. \ref{fig:Natom-pd}. Most
notably, another Mott-type transition occurs within the Mott phase, between
a $x-y$ ferromagnet and a non symmetry breaking Mott state. Related spin ordering
transitions for spin 1 bosons have been discussed recently
by Imembakov {\it et al.} \cite{adilet} and by Snoek and Zhou \cite{snoek}.

The perturbative expansion leading to (\ref{Hduan}) breaks down
as the transition to a superfluid is approached and $t_{a,b}$ become
comparable to $U$.
The question arises, whether the phases predicted
by the effective spin Hamiltonian still hold in this regime. More importantly,
how do the effective spin interactions
affect the nature of the transition
to a superfluid and the superfluid phase itself?

To answer these questions we shall develop in the next two sections a theory which
captures the effective spin interactions while also able to
describe the transition to a superfluid.

\section{Mean field theory of the Superfluid-Mott transition}\label{sec:MFT}

The usual, single component,
Mott transition of bosons is well described by mean field theory \cite{fisher,mott-mft}.
It is thus natural to start our treatment of the two component case with a
mean field approach. In order to capture the superfluid phase we
extend the regime considered in the previous section to allow for arbitrary ratios of
$t_{a,b}/U$. However we shall confine ourselves to the case of a single atom per site
and to the limit $U,t_{a,b}<<V_a,V_b$. Later we shall consider corrections
due to finite intra-species interactions.

In this limit, it is particularly advantageous to use a variational approach,
which is equivalent to mean-field theory \cite{kotliar}. The idea is to
assume a site factorizable wave function associated with hard core bosons, which in our case takes the form
\bea
\ket{\Phi}=&\prod_i&\Bigg[\sin\frac{\t_i}{2}\left(\sin\frac{\x_i}{2}a\yd_i
+\cos\frac{\x_i}{2}b\yd_i\right)\nn\\
&&+\cos\frac{\t_i}{2}
\left(\sin\frac{\h_i}{2}
+\cos\frac{\h_i}{2}a\yd_i b\yd_i\right)\Bigg]\ket{0}.
\label{Phi0}
\eea
The enormous reduction in Hilbert space, made possible by neglecting double
occupation, is what makes these states convenient to work with. Specifically, it is easy
to calculate expectation values. In addition, we shall see that they facilitate
a fluctuation expansion about the mean field theory. Generalization to include
higher occupations is possible but would make the subsequent calculations
much more complicated.
Note that a more general mean-field ansatz would include complex
weights, however it is easily verified that
this would not improve the variational energy.

In the Mott state, where each site is occupied by exactly one atom,
the variational state simplifies even more
\be
\ket{\Phi_{MI}}=\prod_i\left(e^{i\f/2}\sin\frac{\x_i}{2}a\yd_i
+e^{-i\f/2}\cos\frac{\x_i}{2}b\yd_i\right)\ket{0}.
\label{PhiMI}
\ee
It can be viewed as a pseudospin-$\half$ state with $a\yd\ket{0}=\ket{\ua}$
and $b\yd\ket{0}=\ket{\da}$.

\begin{figure}[h]
  \centering
  \includegraphics[width=8cm]{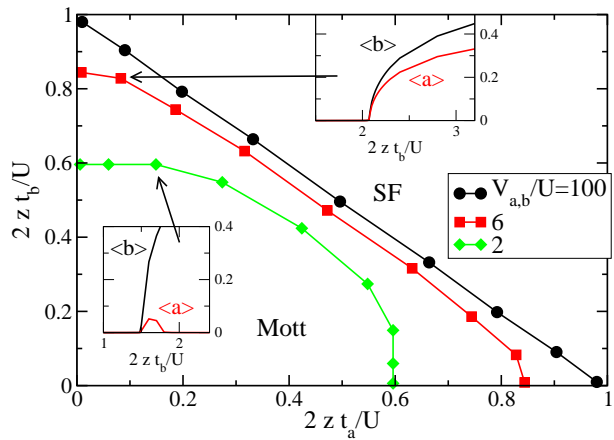}
\caption{Phase diagram obtained by the decoupling mean--field
theory (\protect{\ref{H_{MF}}}) for $U=20$ and different values of
$V_{a,b}$. Note that for finite $V_{a,b}$ and strong asymmetry of
the hopping $t_{a,b}$ one of the species can be completely
depopulated in the superfluid phase (see lower left inset). Except
for this case, the Mott transition always happens at the same
parameters for both species.} \label{fig:mf-pd}
\end{figure}

The onset of superfluidity is characterized by the
development of an order parameter $\sin\t\ne0$.
More precisely, the superfluid order parameters of the two
species in the state $\ket{\Phi}$ are given by:
\bea
\av{a}&=&\half\sin\t\cos\left(\frac{\x-\h}{2}\right)\nn\\
\av{b}&=&\half\sin\t\sin\left(\frac{\x+\h}{2}\right)
\eea
Now a classical energy functional can be written,
which is defined by the expectation value
of (\ref{BHM}) in $\ket{\Phi}$.
Allowing for two sub-lattice order the energy function is:
\bea
E=&&-\frac{z t_a}{4}\sin\t_A\sin\t_B\cos\left(\frac{\x_A-\h_A}{2}\right)
\cos\left(\frac{\x_B-\h_B}{2}\right)\nn\\
&&-\frac{z t_b}{4}\sin\t_A\sin\t_B\sin\left(\frac{\x_A+\h_A}{2}\right)
\sin\left(\frac{\x_B+\h_B}{2}\right)\nn\\
&&+\frac{U}{8}(\cos\t_A+\cos\t_B),
\label{Evar}
\eea
where $z$ is the lattice coordination number.
In the superfluid phase this function
is minimized when both $\cos((\x_i-\h_i)/2)=1$
and $\sin((\x_i-\h_i)/2)=1$, which implies $\x_i=\h_i=\pi/2$.
The remaining degree of freedom $\t$ is uniform on the lattice and
found by minimizing
\be
E(\t)=-\frac{z}{4}(t_a+t_b)\sin^2(\t)+\frac{U}{4}\cos\t
\ee
The result is
\be
\t=\left\{\begin{array}{ll}
\pi & t_a+t_b < t_c \\
\mbox{acos}\left(-t_c/(t_a+t_b)\right) & t_a+t_b> t_c \end{array} \right.
\label{theta}
\ee
where $t_c=U/2z$. We thus
find a transition to a Mott insulating state for $t_a+t_b<t_c$
as illustrated by the circles in Fig. \ref{fig:mf-pd}. This
constitutes a straightforward generalization of the standard
transition for a single species.

By assuming the variational state (\ref{Phi0}), we neglected contributions
from states with multiply occupied $a$ or $b$ bosons. To determine
effects arising from the finite magnitude of the intra--species interaction
we use a numerical self--consistent mean field field theory of (\ref{BHM}).
As first proposed in \cite{mott-mft}, the kinetic energy terms in the
Hamiltonian are decoupled:
\bea
H_{MF} &=& U \sum_i \left(n_{ai} - \frac{1}{2}\right)
\left(n_{bi} - \frac{1}{2}\right) + \frac{1}{2} \sum_{i; \alpha=a,b}
V_\alpha n_{\alpha i} \left(n_{\alpha i} -1\right) \nonumber \\
&& -\sum_{\av{ij}}t_\a\left(a\yd_i \langle a\nd_j\rangle +H.c\right)
-t_b\sum_{\av{ij}}\left(b\yd_i \langle b\nd_j\rangle +H.c\right) + const.
\label{H_{MF}}
\eea
In the homogeneous phase this leads to a sum of identical single--site Hamiltonians
\bea
\tilde{H}_{MF} &=& U \left(n_{a} - \frac{1}{2}\right)
\left(n_{b} - \frac{1}{2}\right) +
\frac{1}{2} \sum_{\alpha=a,b} V_\alpha n_{\alpha} \left(n_{\alpha} -1\right) \nonumber \\
&& - \Psi_a \left(a\yd + H.c. \right)
- \Psi_b \left(b\yd + H.c. \right)
\label{single_site}
\eea
where the \emph{decoupling fields} have to be determined self--consistently according to
\be
\Psi_{a,b} = z t_{a,b} \langle a (b)\rangle.
\label{psi}
\ee
We have solved the combined set of Eqs.~(\ref{single_site}) and (\ref{psi}) numerically
by diagonalizing $\tilde{H}_{MF}$ within a finite--size Hilbert space where we allow for
up to M=9 bosons per species.

We show results in Fig. (\ref{fig:mf-pd}), where it can be seen that
for a small ratio $U/V_{a,b}$ the phase diagram is identical to that determined variationally.
As $V_{a,b}$ decrease and approach $U$ the Mott domain shrinks.
For $V_{a,b}<U$ there is an instability toward a $z$-ferromagnetic superfluid.
Since the experiment is done at fixed magnetization this would lead to
phase separation into domains occupied only by $a$ or by $b$ atoms.
%%%%%%%%%%%%%%%%%%%%%%%%%%%%%%%%%%%

Note that in the Mott state where the order parameters $\av{a}$
and $\av{b}$ vanish, the ground state of $H_{MF}$ has precisely
one atom per site but is completely independent of the relative
weights of $a$ and $b$ atoms. Similarly, the variational energy in
the Mott state (\ref{PhiMI}) is a constant ($-U/4$), independent
of the individual spin orientation. Thus the simple mean field
approaches are unable to resolve spin order in the Mott state. To
obtain spin order we shall in the next section consider
quantum fluctuations around the variational mean field solutions.

\section{Effect of fluctuations: ``magnetic'' states}\label{sec:fluc}

The situation we encountered when attempting to treat the
Mott phase with the variational states is similar to
the basic problem of frustrated quantum magnets.
The classical energy of such systems,
i.e the expectation value of the hamiltonian
in a basis of coherent spin states, often contains a macroscopic degeneracy
(see for example the review Ref. \cite{rev-frus-mag}).
A general mechanism that can lift the degeneracy is ``quantum
order by disorder'', whereby broken symmetry
configurations are selected by the
zero-point energy due to spinwaves \cite{shender}.

A spinwave expansion in magnetism includes the quadratic fluctuations
around coherent-state mean field configurations. We
formulate a similar expansion in fluctuations about the
mean-field states (\ref{Phi0}). As a first step
we define second quantized bosonic
operators, that create the appropriate Hilbert space:
\be
\{a\yd_i\ket{0},b\yd_i\ket{0},a\yd_i b\yd_i\ket{0},\ket{0}\}
\equiv\{\a\yd_{1i}\ket{\W},\a\yd_{2i}\ket{\W},p\yd_i\ket{\W},h\yd_i\ket{\W}\},
\ee
where $\ket{\W}$ is the vacuum of the new bosons and $\ket{0}$ is an empty site.
The new operators
are analogues of Schwinger bosons in spin systems. Like the Schwinger
bosons, they
obey a holonomic constraint, namely that their total filling on a site is one.
Now, we apply an orthogonal change of basis:
\be
\left(\begin{array}{c}
\psi_{0i} \\ \psi\nd_{1i} \\ \psi\nd_{2i} \\ \psi\nd_{3i} \end{array}\right)
=\left(\begin{array}{cccc}
\sin\frac{\t_i}{2}\sin\frac{\x_i}{2} &
\sin\frac{\t_i}{2}\cos\frac{\x_i}{2} &
\cos\frac{\t_i}{2}\sin\frac{\h_i}{2} &
\cos\frac{\t_i}{2}\cos\frac{\h_i}{2} \\
\cos\frac{\t_i}{2}\sin\frac{\x_i}{2} &
\cos\frac{\t_i}{2}\cos\frac{\x_i}{2} &
-\sin\frac{\t_i}{2}\sin\frac{\h_i}{2} &
-\sin\frac{\t_i}{2}\cos\frac{\h_i}{2} \\
\cos\frac{\x_i}{2} & -\sin\frac{\x_i}{2} & 0 & 0\\
0 & 0 & \cos\frac{\h_i}{2} & -\sin\frac{\h_i}{2} \end{array}\right)
\left(\begin{array}{c}
\a\nd_{1i}\\ \a\nd_{2i} \\ p\nd_i \\ h\nd_i \end{array}\right)
\label{rotation}
\ee
In the new basis the variational state (\ref{Phi0})
is simply a singly occupied Fock state of the $\psi\nd_0$ boson
\be
\ket{\Phi}=\prod_i\psi\yd_{0i}\ket{\W}.
\ee
The three remaining bosons, $\psi\yd_{1,2,3}$, create orthogonal
fluctuations about the variational state.

In the constrained Hilbert space of no double occupancy by the same species,
the hamiltonian (\ref{BHM})
may be written in terms of the $\psi$ bosons. Furthermore, $\psi_{0i}$
can be eliminated using the hard core constraint
\be
\psi\yd_{\b i}\psi\nd_{0i}=\psi\yd_{\b i}\sqrt{1
-\sum_{\a=1}^3\psi\yd_{\a i}\psi\nd_{\a i}},
\ee
so that the hamiltonian is a function of only the three fluctuation operators
$\psi_{1,2,3}$. Assuming the fluctuations
are small, we expand it to quadratic order
in these operators. The exact form of the quadratic hamiltonian depends
on the variational starting point which fixes the rotation matrix
(\ref{rotation}). For a two sub-lattice variational state,
the fluctuation hamiltonian has the general form:
\be
H_{fluc}=E_{var}+\half\sum_{\bf k}\left\{\bPsi\yd_k\left(\begin{array}{cc}
\cF_k & {\mathcal G}_k \\
{\mathcal G}^\star_k & \cF^\star_k \end{array}\right)
\bPsi\nd_k-\tr \cF_k\right\}
\label{Hfluc}
\ee
where
\be
\bPsi\yd_\bk=(\bpsi\yd_\bk~~\bpsi\yd_{\bk+\pi}
~~\bpsi^\cT_{-\bk}~~\bpsi^\cT_{-\bk+\pi}), ~~~\bpsi\yd_k=
(\psi\yd_{1\bk}~~\psi\yd_{2\bk}~~\psi\yd_{3\bk})
\ee
while $\cF_k$ and ${\mathcal G}_k$ are $6\times 6$ matrices which depend
on the variational parameters. Finally $H_{fluc}$ is diagonalized
by a Bogoliubov transformation to obtain the excitation frequencies
$\w_{\a\bk}$ and the correction to the ground state energy:
\be
\D E=\half\sum_\bk\left\{-tr\cF_\bk+\sum_\a\w_{\a\bk}\right\}.
\label{DE}
\ee
With the Bogoliubov transformation at hand it should be straightforward
to calculate the average occupation of the fluctuations. For consistency of our
approach we require:
\be
\sum_{\a=1}^3 \langle\psi\yd_{\a i}\psi\nd_{\a i}\rangle<<1
\ee

Let us now focus on the Mott phase. Recall that the
variational energy is independent of the individual spin
orientations, i.e. the parameters in the state (\ref{PhiMI}).
The fluctuation hamiltonian on the other hand will depend on the
spin configuration. Before we compare the zero point energies
corresponding to possible spin orders let us note a few general
properties of the fluctuations in this case.
Since the bosons $p\yd_i$ and $h\yd_i$ which create an extra particle
or hole, are unoccupied in (\ref{PhiMI}), they constitute a pair
of orthogonal fluctuations. The third orthogonal fluctuation
is $\phi\yd=\cos(\x_i/2)\a\yd_{1i}-\sin(\x_i/2)\a\yd_{2i}$ which creates
a pseudospin of opposite orientation. Since the classical energy is
independent of the spin configuration we expect that $\phi\yd_i$ will not
appear in the quadratic fluctuation hamiltonian. This reflects the
fact that a local spin flip does not cost energy.

For a uniform state with $\x_i=\x$ the hamiltonian assumes a simple form
\bea
H_{fluc}&=&\sum_k\Bigg\{f_h(k)h\yd_k h\nd_k+f_p(k)p\yd_k p\nd_k
-\frac{g(k)}{2}(p\yd_k h\yd_{-k}+p\nd_k h\nd_{-k})\Bigg\}
\label{unifluc} \eea
where the couplings depend on $\x$
\bea f_h(k)&=&\frac{U}{2}
-(t_a\cos^2\frac{\x}{2}+t_b\sin^2\frac{\x}{2})z\gk\nn\\
f_p(k)&=&\frac{U}{2}
-(t_a\sin^2\frac{\x}{2}+t_b\cos^2\frac{\x}{2})z\gk\nn\\
g(k)&=&z\gk(t_a+t_b)\sin\x \eea
The hamiltonian is diagonalized by
a standard Bogoliubov transformation:
\bea
p_k&=&\cosh\t_k c\nd_k+\sinh\t_k d\yd_{-k}\nn\\
h_k&=&\cosh\t_k d\nd_k+\sinh\t_k c\yd_{-k}
\label{Bu}
\eea
which yields the excitation modes
\bea
\w_{1,2}(\bk)&=&\half\sqrt{U^2-2U(t_a+t_b)z\gk+(t_a+t_b)^2(z\gk\cos\x)^2}
\pm(t_a-t_b)z\gk \cos\x\nn\\
\label{w_xy}
\eea
In addition, there is a zero mode
$\w_3(\bk)=0$ corresponding to local spin flips, which
reflects the macroscopic degeneracy at the classical level.
Higher order terms in the fluctuations take into account the corrected
potential landscape and generate
a dispersion of the spin flip mode  $\w_3(\bk)$.
Since here we are interested in the zero point energy,
we need not go beyond quadratic fluctuations.
The quantum correction to the ground state energy is
calculated from the prescription (\ref{DE})
\bea
\Delta E(\x)=\frac{1}{2N}
\sum_k\left[\sqrt{U^2-2U(t_a+t_b)z\gk+(t_a+t_b)^2(z\gk \cos\x)^2}
-U+(t_a+t_b)z\gk\right]-\frac{z}{2}\left(\frac{t_a^2}{V_a}+\frac{t_b^2}{V_b}
\right).
\label{DExy}
\eea
where we have added the last term perturbatively in $t_\a/V_\a$.
This is justified in the regime of interest $t_\a,U<<V_\a$.
The minimum of $\D E(\x)$ occurs for $\x=\pi/2$,
which corresponds to pseudospins aligned on the $x-y$ plane.
Note that the dispersions of the particle and hole excitations (\ref{w_xy})
are degenerate in this case. Their gap vanishes when
$t_a+t_b=U/2z$, which marks the transition to a superfluid in agreement
with the variational result (\ref{theta}).

\begin{figure}[h]
  \centering
  \includegraphics[width=8cm]{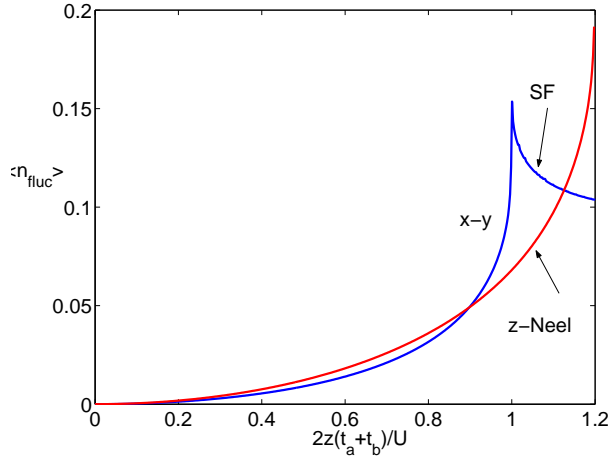}
\caption{Mean square fluctuation in the superfluid and magnetic Mott phases.
The demonstration is at a fixed ratio $t_a/t_b=0.5$.}
\label{fig:fluc}
\end{figure}

To check the consistency of our fluctuation expansion
the local density of fluctuations in the $x-y$ ferromagnet
can be calculated using
the Bogoliubov transformation (\ref{Bu}):
\be
\av{p\yd_i p\nd_i+h\yd_ih\nd_i}=\frac{1}{N}\sum_k 2\sinh 2\t_k
=\frac{1}{N}\sum_k\left(\frac{1-z\gk(t_a+t_b)/U}{\sqrt{1-2z\gk(t_a+t_b)/U}}
-1\right)
\ee
Fig. \ref{fig:fluc} plots the mean square fluctuation as a function of
$(t_a+t_b)/U$ at a constant ratio $t_a/t_b$. It can be verified that
the mean square local fluctuation is smaller than $1/4$ throughout the phase
diagram. This constitutes a posteriori justification for our
expansion which relied on the smallness of the fluctuations.
We should comment though that the occupation of the zero mode cannot be calculated
at this order. If the $x-y$ state is indeed stable, interactions would
generate a dispersion which would lead to a finite local ground state occupation.

We now consider the canted state
\be
\ket{\Psi(\x)}=\prod_{i\in A}\left(\cos\frac{\t}{2}a\yd_i+\sin\frac{\t}{2}b\yd_i\right)
\prod_{i\in B}\left(\sin\frac{\t}{2}a\yd_i+\cos\frac{\t}{2}b\yd_i\right)\ket{0},
\ee
The angle $\t$ parameterizes a continuous path from the
$z$-Neel state ($\t=0$) to the $x-y$ ferromagnet ($\t=\pi/2$).
Since $\ket{\Psi(\t)}$ is not translationally invariant, neither
will be the fluctuation hamiltonian derived from it. An elegant way to
overcome this difficulty is to apply a unitary
particle hole transformation on sub lattice $B$
\bea
\a_{1i}&\leftrightarrow& \a\yd_{2i}\nn\\
p_i&\leftrightarrow& h_i
\label{sublat-rot}
\eea
for $i\in B$.
In the spin language this is equivalent to a $\pi$ rotation of the
spins in the $B$ sub-lattice about their $x$ axis.
The rotation changes the hopping terms in the hamiltonian (\ref{BHM})
\bea
a\yd_i a\nd_j+h.c. \ra a\yd_i a\yd_j +h.c.\nn\\
b\yd_i b\nd_j+h.c. \ra b\yd_i b\yd_j +h.c.,
\eea
but it also transforms $\ket{\Psi(\x)}$ to a translationally invariant state
\be
\ket{\Psi(\x)}\ra\prod_{i}\left(\cos\frac{\t}{2}a\yd_i+\sin\frac{\t}{2}b\yd_i\right)\ket{0}.
\ee
Our procedure can now be carried out with the new hamiltonian and the transformed state.
The fluctuation hamiltonian assumes the form
\bea
H_{fluc}&=&\sum_{\bk}\Bigg\{\frac{U}{2}(p\yd_\bk p\nd_\bk+h\yd_\bk h\nd_\bk)-\frac{z\gk}{2}\sin\t(t_1+t_2)
(p\yd_\bk h\nd_\bk+h.c.)\nn\\
&&-\frac{z\gk}{2}(t_1\cos^2\frac{\t}{2}+t_2\sin^2\frac{\t}{2})(h\yd_\bk h\yd_{-\bk}+h.c.)
-\frac{z\gk}{2}(t_1\sin^2\frac{\t}{2}+t_2\cos^2\frac{\t}{2})(p\yd_\bk p\yd_{-\bk}+h.c.)\Bigg\}
\eea
which can be diagonalized by a Bogoliubov transformation.
In the $z$-Neel state ($\t=0$) the excitation energies assume a particularly simple form
\be
\w_{1,2}(k)=\frac{U}{2}\sqrt{1-\left(\frac{2zt_{a,b}\gk}{U}\right)^2}.
\label{w_z}
\ee
Note that contrary to the $x-y$ state, the
excitations are non degenerate. The gap in $\w_{1,2}(k)$ vanishes
on the lines $t_{a,b}=U/2z$ respectively.
Thus the $z$-Neel state is locally stable toward formation of a
superfluid within these boundaries. There is however a dangerous
zero mode $\w_3(\bk)=0$ which may be destabilized by higher
order terms in the fluctuation hamiltonian. This mode corresponds to $\phi\yd$
which describes spin fluctuations ($\phi\yd_\bk$) toward the
$x-y$ ferromagnetic state. In regions where the
$z$-Neel state is ultimately stable these corrections would just generate
a dispersion for $\w_3(\bk)$.

%%%%%%%%%%%%%%%%%%%%%%%%%%%%%%%%%%%

The quantum zero-point energy of the fluctuations in the $z$-Neel state
is given by:
\be
\Delta E_z=\frac{U}{4}\sum_k\left[
\sqrt{1-\left(\frac{2zt_a\gk}{U}\right)^2}+
\sqrt{1-\left(\frac{2zt_b\gk}{U}\right)^2}-2\right].
\label{DEz}
\ee
The mean local fluctuation can be calculated
in the same way as before
\be
\av{p\yd_i p\nd_i+h\yd_ih\nd_i}=\frac{1}{2N}\sum_k\left[
\frac{1}{\sqrt{1-\left(\frac{2zt_a\gk}{U}\right)^2}}+
\frac{1}{\sqrt{1-\left(\frac{2zt_b\gk}{U}\right)^2}}-2\right].
\ee
It is plotted in Fig. \ref{fig:fluc}, which
demonstrates that fluctuations about the $z$-Neel state are also small.

Before we address the full Mott domain, it is instructive to
evaluate the energy corrections (\ref{DExy}) and (\ref{DEz}) deep
in the Mott phase, where we can compare the result with the
effective spin hamiltonian (\ref{Hduan}).
It is also much easier to evaluate the zero point energy in this limit.
For $t_a,t_b<<U$ we can expand the
square roots in  (\ref{DExy}) and (\ref{DEz}),
then perform the momentum sums exactly, with the result
\bea
\D E_{xy}&\approx&-\frac{z(t_a+t_b)^2}{4U}
-\frac{z}{2}\left(\frac{t_a^2}{V_a}+\frac{t_b^2}{V_b}
\right)\nn\\
\Delta E_z&\approx&-\frac{z(t_a^2+t_b^2)}{2U}
\label{DE_approx}
\eea
These are identical to the mean-field energies in
the effective spin hamiltonian (\ref{Hduan}).
Thus we see that our fluctuation
analysis about the variational states
captures the essential spin interactions. In the next section we
address the stability of the spin states over the entire parameter regime
to derive a phase diagram.

A fluctuation hamiltonian can be derived in a similar way for the superfluid phase
where we find the three excitation modes:
\bea
\w_{1,2}(\bk)&=&\frac{U}{8a}\left[t_+
\pm\sqrt{t_+ +\left(t_-/t_+\right)^2(t_+^2-1)}\right]\bk+O(k^2)\nn\\
\w_{3}(\bk)&=&U\sqrt{t_+^2-1+(\bk/2)^2}
\label{wsf}
\eea
with $t_+=t_a+t_b$ and $t_-=t_a-t_b$.
The zero point energy correction in the superfluid phase
is evaluated using the prescription (\ref{DE}).
A discussion of the collective modes and of the nature of the
superfluid phase is deferred to the next section.

\section{Phase diagram for 1 atom per site}\label{sec:phase-diagram}

In this section we combine the ingredients prepared in the last sections
to present a phase diagram
for a lattice with an average occupation of one atom per site.
From the variational approach we found that the $x-y$ ferromagnet becomes
unstable towards a superfluid state when $t_a+t_b>U/2z$.
The mean field phase diagram was sketched in Fig (\ref{fig:mf-pd}).
However the boundaries of these phases with the
$z$-Neel state remain undetermined.
It is the quantum zero point energy of fluctuations that selects ordered
magnetic states from a degenerate variational energy.

\begin{figure}[h]
  \centering
  \includegraphics[width=8cm]{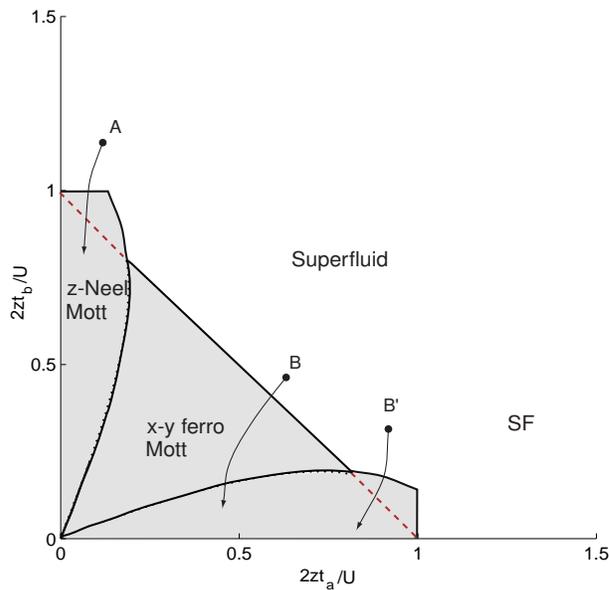}
\caption{Phase diagram including quantum fluctuations.}
\label{fig:pd-fluc}
\end{figure}

To analyze the stability of the spin states
we need to calculate the derivatives
with respect to $\t$ of the zero point energies corresponding to these phases:
\be
\left(\frac{d^n\D E}{d\t^n}\right)_{\t=\t_0}=\sum_{\a,\bk}\left(\frac{d^n \w_\a(\bk)}{d\t^n}\right)_{\t=\t_0}
-\frac{d^n}{d\t^n}\left((t_a^2/V_a+t_b^2/V_b)z\sin^2\frac{\t}{2}\right)_{\t=\t_0}.
\label{dEdn}
\ee
The last term is added perturbatively in $t_\a/V_\a$ and corrects for a large but finite
intra-species interaction. It is easily seen that the first derivative of the modes
$\w_\a$ vanishes identically at the points $\t=0$ and $\t=\pi/2$, corresponding to the $z$-Neel
and $x-y$ states.
Consequently these states are either minima or maxima
of the zero point energy. The second derivative at the $z$-Neel state is given by:
\bea
\left(\frac{d^2\D E}{d\t^2}\right)_{\t=0}=\frac{U/2}{\tau_a-\tau_b}
\sum_\bk\left(\frac{\tau_b(1-\tau_b^2\g_\bk^2)+\tau_a(1+\tau_b^2\gk^2)}{\sqrt{1-\tau_b^2\gk^2}}
-\frac{\tau_a(1-\tau_a^2\g_\bk^2)+\tau_b(1+\tau_a^2\gk^2)}{\sqrt{1-\tau_a^2\gk^2}}\right)
-\frac{U}{2z}\left(\frac{\tau_a^2}{v_a}+\frac{\tau_b^2}{v_b}\right)
\label{d2E}
\eea
where we have denoted $\tau_\a\equiv 2z t_\a/U$ and $v_\a=V_\a/U$.
The domain of stability of the phase is obtained by numerically evaluating
the momentum sum in (\ref{d2E}). The resulting domain of stability is
of the general shape illustrated in Fig. \ref{fig:pd-fluc}. Note that the phase
boundaries deep in the Mott state ($t_{a,b}<<U$) are linear
and coincide with the
result obtained from the effective hamiltonian (\ref{Hduan}).
However we find in contrast with the effective spin hamiltonian,
that even for true hard core interactions $V_{a,b}\ra\infty$, there is a finite
$x-y$ ferromagnetic domain.
%%%%%%%%%%%%%%%%%%%%%%%%%%%%%%%%%%%

\begin{figure}[h]
  \centering
  \includegraphics[width=8cm]{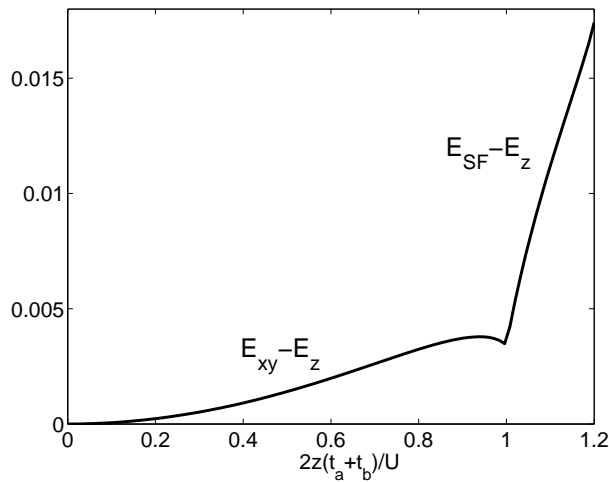}
\caption{The energy, including quantum fluctuations,
of the $z$-Neel state for $t_a/t_b=0.1$.}
\label{fig:DE}
\end{figure}

Note that the Mott $z$-Neel domain in Fig. (\ref{fig:pd-fluc}) extends beyond
the mean field transition to the superfluid which occurs at $t_a+t_b=U/2z$.
As seen in Fig. \ref{fig:DE}, this is due to a lower ground state energy
(including the quantum corrrection) than
the superfluid.
 In the remainder of this section
we shall examine the nature of the phases and transitions in Fig.
\ref{fig:pd-fluc}.
%%%%%%%%%%%%%%%%%%%%%%%%%%%%%%%%%%%

\subsection{Metastability and hysteresis}
It is an interesting observation that over a significant parameter range,
quantum fluctuations favor the $z$-Neel state even where its variational
energy alone is higher than that of the superfluid.
What kind of transition then, is marked by the
lines $t_{a,b}=U/2z$, where the $z$-Neel state finally becomes
unstable? It could be one of the two: (i)
A first order transition into the superfluid state or (ii)
a second order transition
into a supersolid, namely a superfluid that retains Ising order.

We shall see that the former indeed occurs, but this is not immediately obvious.
Consider the excitation modes (\ref{w_z}).
Since only one of them becomes gapless on
the transition lines $t_{a,b}=U/2z$,
one might guess that these lines mark the formation
of a supersolid. However we now show that at the classical level the
supersolid is unstable to formation of a uniform superfluid.

A variational state describing a supersolid is given by
\be
\ket{\Phi_{SS}}=\prod_{i\in A}a\yd_i\left(\sin\frac{\t}{2}+\cos\frac{\t}{2}
b\yd_i\right)\prod_{i\in B}b\yd_i\left(\sin\frac{\t}{2}+\cos\frac{\t}{2}
a\yd_i\right)
\ee
When $\t=\pi$ this is just the $z$-Neel state. A superfluid
component is added to the Neel order for $\t<\pi$. To assess
stability we parameterize, a continuous deformation of this state
toward the uniform superfluid:
\bea
\ket{\Phi(\d)}&=&\prod_{i}\left[\sin\frac{\t}{2}\left(\cos\frac{\x_i}{2}a\yd_i
+\sin\frac{\x_i}{2}b\yd_i\right)+\cos\frac{\t}{2}\left(\sin\frac{\x_i}{2}+
\cos\frac{\x_i}{2}a\yd_i b\yd_i\right)\right]\ket{0}
\eea
where $\x_i=\d,\pi/2-\d$ for $i\in A,B$ and $\d\in[0,\pi/2]$.
The variational energy in this state, calculated using Eq. (\ref{Evar}) is:
\be
E(\t,\d)=-\frac{z(t_a+t_b)}{16}\sin2\t-\frac{zt_a}{16}\sin2\t\sin\d
\ee
The second derivative
$\left(\partial E/\partial\d\right)_{\d=0}$ is negative, indicating that the supersolid
is unstable. We thus establish a first order transition between the $z$-Neel
state and the superfluid. We should point out that at higher order,
quantum fluctuations can the change potential manifold. In particular, they can
make the supersolid phase locally stable. This interesting possibility can
be checked, for example by quantum monte carlo simulations. However, at
our level of approximation there is a first order transition directly
to the uniform superfluid.

An important implication is the
presence of a hysteresis region.
Consider a change of system parameters from the superfluid
to the $z$-Neel state along route A in Fig. \ref{fig:pd-fluc}.
In the region where the superfluid is meta-stable, it may take an excessively long
time to nucleate the $z$-Neel state. The system is thus likely
to remain in the superfluid state until its line of meta-stability
is crossed (dashed lines in Fig \ref{fig:pd-fluc}). Passing
the same route in the opposite direction the $z$-Neel state will of course
persist up to the transition line.
Along routes $B$ and $B'$ matters are qualitatively different.
Since the transition into the $x-y$ Mott state is continuous, there
is no hysteresis there. The transition from the $x-y$ or superfluid to the $z$-Neel state
is first order but without a significant hysteresis region. In fact it
is well described as a transition from easy-axis to easy-plane anisotropy
in the Heisenberg model (\ref{Hduan}).

\subsection{The superfluid and the $x-y$ ferromagnet}
The uniform superfluid phase deserves a closer examination.
In several aspects, it is different
from the single component case.
Most importantly, the two component superfluid is intimately related to the $x-y$ ferromagnet.

In terms of the bosonic operators, the $x-y$ ferromagnetic Mott state sustains
an order parameter $\av{a\yd b}\ne 0$. In this sense
it can be viewed as a counter flow superfluid \cite{kuklov}.
It can support supercurrents of relative
motion characterized by a gradient of the relative
phase between $a$ and $b$ atoms. In the spin language this is
simply a gradient of the spin orientation on the $x-y$ plane.
A magnetic field gradient, $h({\bf x})=h_0 x$ in (\ref{Hduan}),
would twist the spin configuration at a constant rate inducing
AC super-counterflow of frequency $\w=h_0 L$. The goldstone mode associated with this
order is a spinwave, which describes fluctuations of the relative phase
between the two components.
We should note again that in the quadratic fluctuation hamiltonian (\ref{Hfluc}),
spinwaves are dispersionless. This is a direct
consequence of the spin degeneracy at the classical (variational) level.
However, it does not indicate truely vanishing spinwave stiffness. Indeed,
higher order terms in the fluctuations generate a finite
linear spinwave dispersion. This can also be understood from the
effective spin hamiltonian (\ref{Hduan}), which obviously has a finite
spin stiffness $\propto J_\perp$.

As system parameters are varied across the transition at $t_a+t_b=U/2z$,
one of the gapped particle-hole fluctuations of the $x-y$ Mott state condenses,
marking the formation of the two superfluid order parameters $\av{a}$ and $\av{b}$.
As pointed out in ref. \cite{kuklov-xy}, in the $x-y$ Mott state only the relative
phase is fixed while the average phase of the two components is disordered. In the
superfluid the average phase orders as well. Accordingly we find
two linear gapless modes in the superfluid.
(\ref{wsf}), corresponding to an in-phase fluctuation of the two components
($\w_1(\bk)$), and a relative phase, spinwave fluctuation $\w_2(\bk)$.
From these considerations it is obvious that the universal aspects of the
transition will be identical to the standard, single component Mott transition \cite{kuklov-xy}.

\section{Discussion and conclusions}

It is interesting to consider the present results in light of the current experimental possibilities.
First of all, we have shown that magnetic phases are robust in the sense  that they persist up to
the transition into the superfluid state.
Specifically, spin-ordered states appear even near the boundary of this phase
transition.  Spin-exchange interactions in
this regime are most
easily accessible experimentally since the relevant
energy scales  are largest and comparable to on-site interaction. In this regime magnetic phases are
therefore relatively insensitive to perturbations due to e.g. inhomogeneous magnetic field variations.
Secondly, we note the existence of several
metastable states in this regime, indicating that the system is likely to display interesting dynamics as the
optical potential is lowered across the transition point. In particular, hysteresis and abrupt changes in
the state of the system can be expected. At the same time, our results indicate that spin-ordered states are
qualitatively different for odd and even numbers of particles per site. Both are likely to be observable  in
any realistic realization, since the inhomogeneous trapping potential typically
leads to domains with different
occupation.

Finally, it is important to note that detection of  the complex states, of the type discussed in this paper,
presents an interesting challenge in its own right.  It turns out that the quantum nature
of strongly correlated magnetic states can be revealed
by {\em spatial noise} correlations in the image of the expanding gas\cite{ehud-noise}.
Specifically, atoms released
from a Mott-insulating state of the optical lattice display sharp (Bragg) peaks in the
density-density correlation function as a
consequence of quantum statistics and such  peaks
can be used to probe the spin ordered Mott states proposed for
two component bosons.

In summary, we presented a theoretical analysis of the phase diagram of
two component bosons on an optical lattice. We extended earlier treatments
which were valid only deep in the Mott phase toward the MI-SF transition and beyond
and were thus able to map a complete phase diagram.
In addition we identified a transition into a Mott phase with no broken
symmetries, which occurs only at even fillings.

\section{Acknowledgments}
We would like to thank L. Mathey, L. Duan, A. Imembakov, I. Bloch, A. Sorensen and D. W. Wang
for useful discussions.  This work was supported by
ARO, NSF (PHY-0134776, DMR-0132874), Sloan Foundation, Packard Foundation, and
the German Science Foundation (DFG).

\end{document}